\documentclass[floatfix,showpacs,aps,nofootinbib,twocolumn]{revtex4}
\usepackage{amssymb,amsmath}
\usepackage{graphicx,float}
\usepackage{indentfirst}
\usepackage{hyperref}
\newcommand{\beq}{\begin{equation}}
\newcommand{\eeq}{\end{equation}}
\newcommand{\bea}{\begin{eqnarray}}
\newcommand{\eea}{\end{eqnarray}}

\def\({\left(}
\def\){\right)}

\begin{document}

\title{The local vicinity of spins sum for certain mass dimension one spinors}
\author{R. J. Bueno Rog\'erio}
\email{rodolforogerio@feg.unesp.br}
\author{J. M. Hoff da Silva}
\email{hoff@feg.unesp.br}
\affiliation{Departamento de F\'{\i}sica e Qu\'{\i}mica, Universidade Estadual Paulista,
Guaratinguet\'{a}, SP, Brazil.}
\begin{abstract}
It is well known that the usual formulation of Elko spinor fields leads to a subtle Lorentz symmetry break encoded in the spin sums. Recently it was proposed a redefinition in the dual structure, along with a given mathematical device, which eliminate the Lorentz breaking term in the spin sums. In this work we delve into the analysis of this mathematical device providing a solid framework to the used method.  
\end{abstract}

\maketitle

\section{Introduction}
Quite recently it was proposed \cite{AH1} a spin one-half fermionic field endowed with canonical mass dimension one which is local, in contrary of what happens with the previous formulations \cite{AH2,AH3}. The whole construction rests upon a remaining degree of freedom in the dual definition. It is believed that this freedom allows to circumvent the rigidity of Weinberg's no-go theorem concerning spin one-half fermions \cite{WEI}. 

The new dual possibility can be obtained from a two steps program: firstly, working out a non-vanishing real Lorentz invariant associated norm. Secondly, evaluating the spin sums and imposing Lorentz covariance in the final result. In order to accomplish the second step two more requirements are necessary: a redefinition on the dual structure, keeping all the previous physical contents, and what is being called the spin sums $\tau-$deformation \cite{AH1,CYL}. After all, these procedures lead to a local fermionic field. 

The aim of this paper is to formalize the last step, by means of the concept of the vicinity of singular matrix. We also delve into the failure of usual methods as the Moore-Penrose pseudoinverse \cite{MOO,PEN} (or generalized inverse) and the Tikhonov's regularization \cite{TIK}. This work is organized as follow: in the next section we review the basic points in the derivation of the new dual. Going forward, in Section III we explicit the problem concerning the Lorentz covariance of the spin sums and formalize the procedure performed in Ref. \cite{AH1}, putting the new spin sums in a robust context. In the final section we conclude.   

\section{The dual}
There is a judicious program leading to an adequate dual associated to a given spinor field \cite{AHPLB}. For book keeping purposes we shall depict the main aspects of this formulation. Let $\psi$ be a given spinor field carrying a $(1/2,0)\oplus(0,1/2)$ representation. Suppose its dual given by $\tilde{\psi}$. In order to reach the appropriate dual, consider the following general form
\begin{equation}
\tilde{\psi_h}(p^\mu)=[\Delta \psi_h(p^\mu)]^{\dagger}\eta, \label{1}
\end{equation} where $h$ stands for the helicity, and impose a real Lorentz invariant non-null norm. The $\Delta$ operator must obey a set of requirements. Denoting the spinor space by $S$, then the $\Delta$ operator is such that 
\begin{eqnarray}
\Delta: S&\rightarrow& S \nonumber\\ \psi_h&\mapsto& \psi_{h'}. \label{2}
\end{eqnarray} Besides, $\Delta$ has to be idempotent ensuring an invertible mapping. Lorentz invariance imply $\eta\sim \gamma_0$, and from Eq. (2) we may have two possibilities: $h=h'$, for which $\Delta=\mathbb{I}$ and it stands for the Dirac usual case, or $h\neq h'$ leading to a more involved operator, which may be represented by \cite{AH1} $\Delta=m^{-1}G(\phi)\gamma_\mu p^{\mu}$, being $m$ the rest mass and $p^\mu$ the four momenta associated to the particle to be described by the quantum field. Moreover, the $G(\phi)$ matrix reads  
\begin{eqnarray}\label{gphi}
\mathcal{G}(\phi)= \left(\begin{array}{cccc}
0 & 0 & 0 & -ie^{-i\phi} \\ 
0 & 0 & ie^{i\phi} & 0 \\ 
0 & -ie^{-i\phi} & 0 & 0 \\ 
ie^{i\phi} & 0 & 0 & 0
\end{array}  \right),
\end{eqnarray} where $\phi$ is the polar angle in the momentum parameterization $p^{\mu}=(E,p\sin\theta\cos\phi,p\sin\theta\sin\phi,p\cos\theta)$. Finally, the formal structure of the spinor itself must be evinced. Hence, by keeping in mind the desired canonical mass dimension, and the neutrality of the resulting field, it is possible to see that the spinor field must be an eigenspinor of the charge conjugation operator \cite{AH1,AH2} $C\psi^{S/A}(p^\mu)=\pm \psi^{S/A}(p^\mu)$, where $\psi^S$($\psi^A$) means (anti)self-conjugated spinor.  

The aforementioned reasoning leads, after a bit of calculation, to the spin following spin sums
\begin{eqnarray}
\sum_h \psi^{S/A}_h(p^\mu)\tilde{\psi}^{S/A}_h(p^\mu)=\pm m[\mathbb{I}\pm G(\phi)],\label{4}
\end{eqnarray} revealing a subtle break in the Lorentz covariance. Here is where the usual Elko construction stops. In the next section we shall pursue the dual redefinition and accomplish the formal $\tau-$deformation. 

\section{The spin sums inverse}
An attempt to solve the Elko breaking Lorentz covariance that emerges in the spin sums, is given by a redefinition in the dual structure, as it can be seen in Ref. \cite{AH1}. Such redefinition reads
\begin{eqnarray}
\tilde{\psi}^{S}_h(p^\mu) &\rightarrow& \tilde{\psi}^{S}_h(p^\mu)\mathcal{A}, \\
\tilde{\psi}^{A}_h(p^\mu) &\rightarrow &\tilde{\psi}^{A}_h(p^\mu)\mathcal{B},
\end{eqnarray}
where the operators $\mathcal{A}$ and $\mathcal{B}$ demand to have some important properties: the spinors $\psi^{S}_h(p^\mu)$ and $\psi^{A}_h(p^\mu)$ must to be eigenspinors of $\mathcal{A}$ and $\mathcal{B}$ respectively, with eigenvalues given by the unity
\begin{eqnarray}\label{7}
\mathcal{A}\psi^{S}_h(p^\mu) = \psi^{S}_h(p^\mu), \quad\quad \mathcal{B}\psi^{A}_h(p^\mu)=\psi^{A}_h(p^\mu).
\end{eqnarray} 
Besides, such operators must to fulfill 
\begin{eqnarray}\label{8}
\tilde{\psi}^{S}_h(p^\mu)\mathcal{A}\psi^{A}_h(p^\mu) = 0, \quad\quad \tilde{\psi}^{A}_h(p^\mu)\mathcal{B}\psi^{S}_h(p^\mu) = 0.
\end{eqnarray}
The set of equations \eqref{7} and \eqref{8} ensures the accuracy of the orthonormality relations, as remarked in \cite{AH2,AH3}, to remain unchanged. With the new dual structure in mind one can evaluate the spin sums, which now will carry  the operators $\mathcal{A}$ and $\mathcal{B}$. A direct calculation leads to
\begin{eqnarray}\label{spinsumab}
\sum_h \psi^{S}_h(p^\mu)\tilde{\psi}^{S}_h(p^\mu)=m[\mathbb{I}+G(\phi)]\mathcal{A}, \\
\sum_h \psi^{A}_h(p^\mu)\tilde{\psi}^{A}_h(p^\mu)=-m[\mathbb{I}-G(\phi)]\mathcal{B}.
\end{eqnarray}

Now, in order to acquire Lorentz covariant spin sums, it could be imposed that $\mathcal{A}$ and $\mathcal{B}$ are simply the inverse of $[\mathbb{I}+G(\phi)]$ and $[\mathbb{I}-G(\phi)]$ respectively \cite{AH1}. However, $\det[\mathbb{I}\pm G(\phi)]=0$ and this naive approach does not work. An interesting general inverse was introduced in the literature by the outstanding work of Moore \cite{MOO} and further developed by Penrose \cite{PEN}. In general grounds these works give a complete algorithm to find out the so-called pseudoinverse, hereafter denoted by $M^{+}$, of a singular matrix $M$. 

As already noted in Ref. \cite{CYL}, unfortunately this procedure renders to be innocuous. In fact, by applying the pseudoinverse protocol, the $[\mathbb{I}+\mathcal{G}(\phi)]^{+}$ is given by \cite{PEN} 
\begin{eqnarray}
[\mathbb{I}+\mathcal{G}(\phi)]^{+} = [\mathbb{I}+\mathcal{G}(\phi)]^{\dagger}\bigg([\mathbb{I}+\mathcal{G}(\phi)][\mathbb{I}+\mathcal{G}(\phi)]^{\dagger}\bigg)^{-1}\!\!, 
\end{eqnarray}
or, in a similar fashion
\begin{eqnarray}
[\mathbb{I}+\mathcal{G}(\phi)]^{+} = \bigg([\mathbb{I}+\mathcal{G}(\phi)]^{\dag}[\mathbb{I}+\mathcal{G}(\phi)]\bigg)^{-1}[\mathbb{I}+\mathcal{G}(\phi)]^{\dag}\!.
\end{eqnarray} Nevertheless, both resulting amounts to be also singular, invalidating this approach. It is necessary, then, to look for an alternative method.

The method known as the Tikhonov's Regularization \cite{TIK} provides another way to calculate the inverse of a singular matrix, given the failure of the previous formulation. Adapting this method to the problem at hand, one is lead to 
\begin{eqnarray}
[\mathbb{I}+\mathcal{G}(\phi)]^{+}\!\!=\!\!\lim_{\tau\rightarrow 0}[\mathbb{I}+\mathcal{G}(\phi)]^{\dag}\bigg([\mathbb{I}+\mathcal{G}(\phi)][\mathbb{I}+\mathcal{G}(\phi)]^{\dag} +\mu\mathbb{I}\bigg)^{-1}\!\!,
\nonumber
\end{eqnarray} where $\mu \in\mathbb{C}$ is a deformation-like parameter. A fairly trivial calculation allows one to recast the above equation as
\begin{eqnarray}\label{14}
[\mathbb{I}+\mathcal{G}(\phi)]^{+} = \lim_{\mu\rightarrow 0} \left(\begin{array}{cccc}
\frac{1}{4+\mu} & 0 & 0 & \frac{-ie^{-i\phi}}{4+\mu} \\ 
0 & \frac{1}{4+\mu} & \frac{ie^{i\phi}}{4+\mu} & 0 \\ 
0 & \frac{-ie^{-i\phi}}{4+\mu} & \frac{1}{4+\mu} & 0 \\ 
\frac{ie^{i\phi}}{4+\mu} & 0 & 0 & \frac{1}{4+\mu}
\end{array}  \right).
\end{eqnarray} In order to verify if \eqref{14} matches the pseudoinverse of $[\mathbb{I}+G(\phi)]$, we multiply $[\mathbb{I}+\mathcal{G}(\phi)]^{+}$ by $[\mathbb{I}+\mathcal{G}(\phi)]$ and take the limit $\mu\rightarrow 0$. As it can be readily verified, the result is 
\begin{eqnarray}
[\mathbb{I}+\mathcal{G}(\phi)][\mathbb{I}+\mathcal{G}(\phi)]^{+} &=& \frac{1}{2}\left(\begin{array}{cccc}
1 & 0 & 0 & -ie^{-i\phi} \\ 
0 & 1 & ie^{i\phi} & 0 \\ 
0 & -ie^{-i\phi} & 1 & 0 \\ 
ie^{i\phi} & 0 & 0 & 1
\end{array}  \right)\nonumber\\
&=&\frac{1}{2}[\mathbb{I}+\mathcal{G}(\phi)], \label{qualquer}
\end{eqnarray} leading, again, to an inefficient method. 

Aiming to find a matrix that can be really classified as an inverse to the Lorentz break part of the spin sums, use was made of a ``$\tau-$deformation'' \cite{AH1}, writing 
\begin{eqnarray}\label{spinsumab2}
\sum_h \psi^{S}_h(p^\mu)\tilde{\psi}^{S}_h(p^\mu)&=&m[\mathbb{I}+\tau G(\phi)]\mathcal{A}\vert_{\tau\rightarrow 1},
\end{eqnarray} and 
\begin{eqnarray}\label{outra}
\sum_h \psi^{A}_h(p^\mu)\tilde{\psi}^{A}_h(p^\mu)&=&-m[\mathbb{I}-\tau G(\phi)]\mathcal{B}\vert_{\tau\rightarrow 1}.
\end{eqnarray} Notice that in order to the $\tau-$deformation makes sense it must have a well defined $\tau\rightarrow 1$ limit. With effect, this limit is the unique necessary constraint used. The investigation of the vicinity of $[\mathbb{I}+\mathcal{G}(\phi)]$ \cite{barata} corroborates this approach, giving full sense to the adopted inverse.  

We start from the fact that both matrices $\mathbb{I}$ and $\mathcal{G}(\phi)$ are non-singular\footnote{We remark, parenthetically, that the very same procedure and conclusions can be {\it mutatis mutandis} applied to the $[\mathbb{I}-G(\phi)]$ case.}. Hence it is possible to see that 
\begin{eqnarray}
\mathbb{I}+\tau G(\phi)=[\mathbb{I}\tau+G^{-1}(\phi)]G(\phi),\label{qq1}
\end{eqnarray} in such a way that $[\mathbb{I}+\tau G(\phi)]$ is invertible as far as $[\mathbb{I}\tau+G^{-1}(\phi)]$ be non-singular. Now take $Z=-G^{-1}(\phi)$ and let $p_z$ be the associated polinomial of $Z$, i. e., $p_z=\det[Z-\lambda\mathbb{I}]$. Notice from Eq. (\ref{qq1}) that 
\begin{eqnarray}
\mathbb{I}+\tau G(\phi)=-[Z-\mathbb{I}\tau]G(\phi),\label{qq2}
\end{eqnarray} and therefore if $\tau$ does not belong to the set of roots of $p_z$, $\{\lambda_k\}$ ($p_z(\lambda_k)=0$), then $[\mathbb{I}+\tau G(\phi)]$ is invertible. Thus, by taking $r_1\equiv\min\{|\lambda_k|,\lambda_k\neq0\}$ and $r_2\equiv\max\{|\lambda_k|\}$ we have two distincts ranges of values for $\tau$ in which 
$[\mathbb{I}+\tau G(\phi)]$ is invertible, namely $\{\tau\in\mathbb{R},0<|\tau|<r_1\}$ and $\{\tau\in\mathbb{R},|\tau|>r_2\}$. As it can be seen 
\begin{equation}
p_z=\lambda^4-2\lambda^2+1,\label{qq3}
\end{equation} whose roots are given by $\pm 1$, both with multiplicity two. Thus, one can see that the unique constraint is given by $|\tau|>1$ or $0<|\tau|<1$. Therefore the limit used in Ref. \cite{AH1} is valid and the operators $\mathcal{A}$ and $\mathcal{B}$ may well be chosen in order to give the inverses of (\ref{spinsumab2}) and (\ref{outra}), respectively.  

\section{Final Remarks}
We have studied some well defined methods, aiming to find the inverse of a singular matrix emerging in the spin sums related to Elko spinors usual formulation. After to try the standard methods encoded in the Moore-Penrose pseudoinverse and the Tikhonov's regularization, we turned to the study of formalizations concerning the neighborhood of singular matrices. Within this context, we found that the $\tau-$deformation provides an acceptable inverse (also) in the limit $\tau\rightarrow 1$. It turns out that this same limit is necessary in the $\tau-$deformation used in Ref. \cite{AH1}. Therefore, the study of the vicinity of the Elko spin sums places the method used in \cite{AH1} in a well posed mathematical level. 

\section{Acknowledgments}
JMHS would like to thanks CNPq (445385/2014-6; 304629/2015-4) for partial support.

\end{document}